\documentclass[aps,prb,twocolumn,superscriptaddress,showpacs]{revtex4}

\usepackage{graphicx}
\usepackage{amsfonts,amsmath,amssymb,amsthm}

\def\be{\begin{equation}}
\def\ee{\end{equation}}
\def\ba{\begin{eqnarray}}
\def\ea{\end{eqnarray}}

\def\bc{\begin{center}}
\def\ec{\end{center}}

\begin{document}

\title{Optical manifestation of the Stoner ferromagnetic transition in 2D electron systems}

\author{A.~B.~Van'kov}
\affiliation{Institute of Solid State Physics, RAS, Chernogolovka
142432, Russia}\affiliation{National Research University Higher
School of Economics, Moscow, 101000, Russia}

\author{B.~D.~Kaysin}
\affiliation{Institute of Solid State Physics, RAS, Chernogolovka
142432, Russia}

\author{I.~V.~Kukushkin}
\affiliation{Institute of Solid State Physics, RAS, Chernogolovka
142432, Russia}\affiliation{National Research University Higher
School of Economics, Moscow, 101000, Russia}

\date{\today}

\begin{abstract}
We perform a magneto-optical study of a two-dimensional electron
systems (2DES) in the regime of the Stoner ferromagnetic
instability for even quantum Hall filling factors on
Mg$_x$Zn$_{1-x}$O/ZnO heterostructures. Under conditions of
Landau-level crossing, caused by enhanced spin susceptibility in
combination with the tilting of the magnetic field, the transition
between two rivaling phases\textemdash paramagnetic and
ferromagnetic\textemdash is traced in terms of optical spectra
reconstruction. Synchronous sharp transformations are observed
both in the photoluminescence structure and parameters of
collective excitations upon transition from paramagnetic to
ferromagnetic ordering. Based on these measurements, a phase
diagram is constructed in terms of the 2D electron density and
tilt angle of the magnetic field. Apart from stable paramagnetic
and ferromagnetic phases, an instability region  is found at
intermediate parameters with the Stoner transition occurring at
$\nu\approx 2$. The spin configuration in all cases is
unambiguously determined by means of inelastic light scattering by
spin-sensitive collective excitations. One indicator of the spin
ordering is the intra-Landau-level spin exciton, which acquires a
large spectral weight in the ferromagnetic phases. The
other\textemdash is an abrupt energy shift of the intersubband
charge density excitation due to change in the many-particle
energy contribution upon spin rearrangement. From our analysis of
photoluminescence and light scattering data, we estimate the ratio
of surface areas occupied by the domains of the two phases in the
vicinity of a transition point. In addition, the thermal smearing
of a phase transition is characterized.
\end{abstract}


\maketitle

\section{Introduction}

Two-dimensional electron systems (2DES) with strong interaction
have attracted considerable research attention from the
perspective of fundamental physics. The technological perfection
of novel 2D materials extends their parameter space far beyond the
framework of GaAs-based heterostructures. Interesting examples
include 2D Fermi liquids in ZnO-based heterostructures, which
exhibit an unusual combination of a large value of the interaction
parameter $r_s$ with ultrahigh electron mobility
level~\cite{FalsonGrowth,ReviewZnO}. New correlated states in
magnetic fields may appear as a result of the interplay between
key energy scales: cyclotron, Zeeman splitting, and inter-particle
Coulomb energy. The material parameters of ZnO-based structures
with 2DES facilitate close proportion between energy scales in
magnetic fields: Zeeman splitting ($g^*_{bulk} \sim 2$) and
cyclotron energy ($m^*_{bulk}\sim 0.28\,m_0$). At moderate
magnetic fields, the Coulomb energy has a substantially higher
scale, thus resulting in high values of the Landau level (LL)
mixing parameter and entangling the familiar single-particle
energy spectrum. Nevertheless, the energy level sequence may be
partially restored if renormalization of Fermi-liquid parameters
is taken into account. In this context, in previous works, a
substantial renormalization of the electron effective mass and
spin susceptibility has been observed in ZnO at large values of
the interaction parameter
$r_s$.~\cite{Stoner2008,Stoner2012,Maryenko2014} As a result,
opposing spin levels of adjacent LLs may approach closer to each
other in terms of energy than those of the same LL. If they happen
to intersect, spontaneous symmetry breaking between two rivaling
spin configurations is possible. At integer Landau level fillings,
the corresponding phases are called quantum Hall ferromagnets
(QHFs) and can be treated within the confines of the Ising model.
Experimentally, this Ising ferromagnetic (FM) transition is
triggered either by tuning the electron's spin susceptibility, or
if the spin splitting is not sufficiently large, by tilting the
external magnetic field at definite {\it coincidence} angles.

The appropriate coincidence angles can be estimated from the
following simplified single-particle relation:
\begin{equation}\label{Eq1}
\frac{E_z}{\hbar \omega_c}=\frac{g^*m^*}{2\,cos\,\Theta}=j,
\end{equation}
where $g^*m^*$ represents the effective spin susceptibility,
$\Theta$ the tilt angle of the magnetic field, and integer $j$ the
coincidence index. This formula has been regularly utilized in
magnetotransport experiments for probing the spin susceptibility,
but it lacks accuracy due to the influence of the many-particle
interaction on the spin splitting and effective mass. Its
applicability at small integer or fractional filling factors is
violated due to the strong exchange effects, which lead to
modification of the energy spectrum ~\cite{Maryenko2014,
FalsonFQHE2015}. The physics of fractional quantum Hall states may
be also modified by level crossing, though in this case levels of
composite fermions play the role. In particular, certain unknown
even-denominator fractional states have been detected in ZnO-based
2DES~\cite{FalsonFQHE2015} at tilt angles different from
single-particle values obtained from Eq.(\ref{Eq1}).

The physics of Ising QHFs at small integer filling factors is
essentially a many-body problem, since it is governed by the
exchange interaction. As has been established earlier in a series
of magnetotransport studies of 2DESs in different heterostructures
~\cite{AlAs2000,AlAs2003,IsingPolish2002,IsingInSb}, a Stoner
ferromagnetic transition at integer filling factors involves the
formation of domains. The transition point itself was identified
by the appearance of sharp resistance spikes, although important
physical parameters such as the energy spectrum of the different
phases, areas occupied by them, and the stability of domains
remained unknown.

In this context, here, we present the first optical study of a
Stoner ferromagnetic transition in a set of ZnO-based
he\-te\-ro\-struc\-tu\-res with 2DESs in the regime corresponding
to the integer quantum Hall effect. Transformations of the $\nu=2$
ground state are carefully studied, and consonant events for
filling factors of 3, 4 and 6 are also traced. The conditions for
a Stoner transition are determined in view of abrupt
transformations in the 2DES energy spectra. Qualitative
reconstruction is detected both in the magneto-photoluminescence
(magneto-PL) spectrum and parameters of the electronic collective
excitations, as probed by inelastic light scattering (or the
Raman) technique. The phase diagram of Ising QH ferromagnets at
$\nu=2$ is acquired in terms of the critical tilt angle as a
function of the electron density. Three essential regions are
identified in the diagrams: stable paramagnetic (PM) region, FM
region and an instability region, wherein the Stoner transition
takes place at $\nu\approx 2$. At electron densities of
$n_{s}<2\times 10^{11}$\,cm$^{-2}$, FM order is observed to
spontaneously develop even at normal orientation of the magnetic
field. From an analysis of the PL and Raman spectra, the ratio of
areas occupied by the domains of the two rivaling phases across
the transition point is estimated. Furthermore, the domain sizes
and their thermal stability are probed.

\section{Experimental technique}
%What is the role of exchange contribution to the transition point.

Measurements were performed on a series of Mg$_x$Zn$_{1-x}$O/ZnO
heterostructures grown by liquid-ozone-assisted molecular beam
epitaxy\cite{FalsonGrowth}. Each structure contained a
high-quality 2D electron channel of varying density, as defined by
the Mg-content in the barrier. Key parameters of the as-grown
samples were characterized by magnetotransport analysis and are
summarized in Table I.

\begin{table}
\caption{ Parameters of two-dimensional electron system (2DES) in
the set of studied samples in order of increasing electron
density. The electron density $n_s$ was measured using the
magne\-to-\-pho\-to\-lu\-minescence technique. Mobility $\mu_{t}$
was qualified by magnetotransport measurements.} \label{table1}

\begin{ruledtabular}
\begin{tabular}{cccc}

Sample ID  & $n_{s}$ ($10^{11}$ cm$^{-2}$) & $\mu_{t}$ ($10^{3}$ cm$^2/$V$\cdot$s)\\
\hline
254 & 1.14 & 710\\
259 & 1.8 & 570\\
244 & 2.3 & 400\\
427 & 2.8 & 427\\
426 & 3.5 & 410\\
448 & 4.5 & 250\\

\end{tabular}
\end{ruledtabular}
\end{table}

Experiments were conducted at low temperatures in the range of
$0.3-4.2\ {\rm K}$ using the He$^3$ evaporation inset to the
cryostat with a superconducting solenoid. Samples were mounted on
a rotational stage in order to control their orientation with
respect to the magnetic field direction. Tilt angles were tuned
{\it in situ} with discrete steps with a finesse of
$\sim0.5^{\circ}$. The applied magnetic fields spanned the range
of 0 to 15\,T.

Optical access to the sample was established via two quartz
fibers, one of which was used for pho\-to\-exci\-ta\-tion, while
the other was used for signal collection. This optical scheme
benefits from a higher signal output and the absence of background
parasitic scattering from the fiber core. The angle configuration
of the fibers determines the momentum transferred from light to
the 2DES. Photoexcitation was produced by a tunable laser source
operating in the vicinity of the direct interband optical
transitions of ZnO. The laser source was designed as a
frequency-doubled tunable continuous-wave Ti-sapphire laser with
output monochromatic radiation in the wavelength range of 365 to
368\,nm. A barium borate crystal set in the single-pass
configuration was utilized as a nonlinear element. The typical UV
excitation power was 2\textendash7\,$\mu$W, which was distributed
over an excitation spot with a surface area $\sim 1\,{\rm mm}^2$.
Thus, the excitation power density was well below 1\,${\rm
mW}/{\rm cm}^{2}$, and which prevented the heating of 2D
electrons. Optical spectra were detected with the use of a
spectrometer in conjunction with a liquid-nitrogen-cooled CCD
-camera. In the UV range, the system exhibited a linear dispersion
of $5\ {\rm \AA/mm}$ and spectral resolution of $0.2\ {\rm \AA}$.

The dynamics of each sample was extensively studiedwith respect to
the magnetic-field orientation in order to study the recombination
spectrum transformation in the vicinity of integer filling
factors. In addition, PL was utilized for electron density
characterization, as described previously~\cite{APL2015}, and for
determining the resonant conditions of inelastic light scattering.

\begin{figure}[htb!]
\includegraphics[width=.48\textwidth]{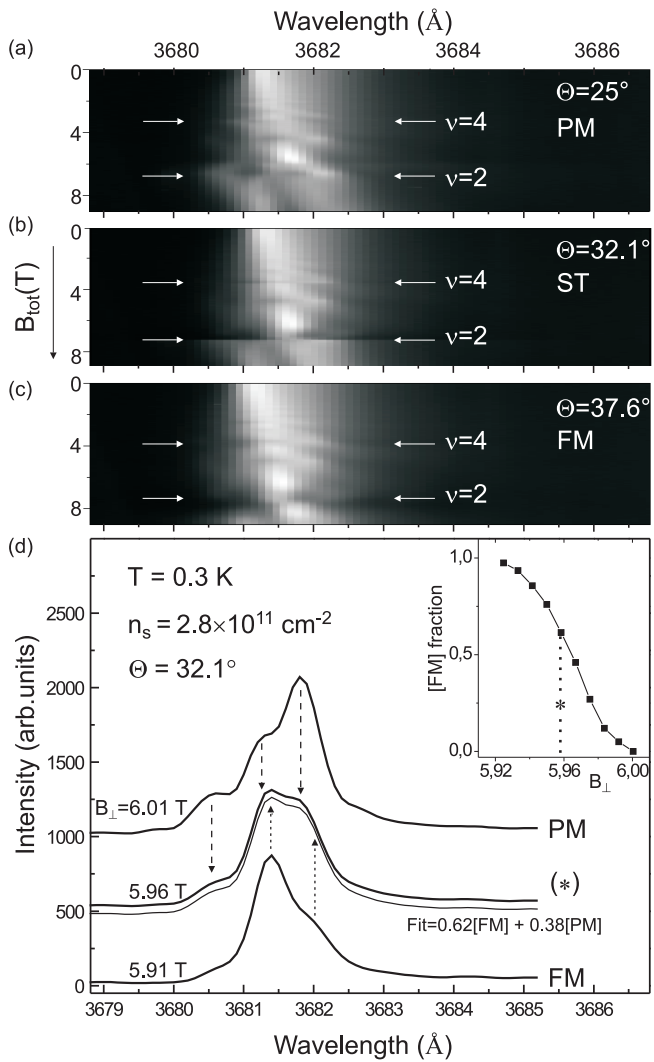}
\vspace{-2.mm} \caption{(a), (b), and (c) -- Image plots of the
photoluminescence (PL) magnetic-field dynamics of the 2DES in
MgZnO/ZnO heterostructure with $n_s=2.8\times10^{11}\,{\rm
cm}^{-2}$ measured at three sample orientations $\Theta$,
corresponding to qualitatively different behaviors at even filling
factors: (a) paramagnetic, (b) Stoner transition, and (c)
ferromagnetic ordering. Positions with distinctive PL behavior at
$\nu=2$ and 4 are indicated by double horizontal arrows. In panel
(b), cutoff positions are discernible by image discontinuity. (d)
Modification of PL spectra around $\nu=2$ at ferromagnetic and
paramagnetic phases and their superposition at a Stoner transition
point (cutoff field). In all cases, the temperature of the sample
was close to 0.3\,K. The inset illustrates a fraction of the
ferromagnetic phase as a function of the magnetic field around the
transition point, as extracted from the superposition fit. The
asterisk marks the position of the cutoff field.} \label{fig3}
\vspace{-1.5mm}
\end{figure}

Inelastic light scattering signal was studied predominantly at
magnetic fields corresponding to LL filling factors of $\nu=2$,
$\nu=1$, and values in between. The multitude of spectral lines
was dominated by a PL signal residing at stationary wavelength
positions, and it was not affected by laser tuning. Raman lines,
although weak, could be distinguished by their constant energy
shift from the sweeping laser position. An important observation
here is that the intensities of the Raman features from 2D
electrons behave resonantly while crossing definite PL bands. This
trick has been thoroughly described in the previous paper, devoted
to collective excitations in ZnO-based 2DESs at zero magnetic
field~\cite{RamanPRB2016}. Here, at high magnetic fields, the
Raman line search procedure was essentially the same as in the
previous study, and only the resonant contours were slightly
shifted by the applied magnetic field.

\section{Optical response of quantum Hall ferromagnets}

The first indication of the rearrangement in LLs is manifested in
the modification of the magneto-photoluminescence dynamics.
Figures 1 a-c present the PL evolution of the 2DES in Sample 427
(density $n_s=2.8\times 10^{11}$\,cm$^{-2}$) as image plots with
the magnetic field increasing along the downward y-direction. The
dynamics shown in panel (a) ($\Theta=25^{\circ}$) and those at
smaller angles are all nearly equivalent if corrected to the
normal component of the magnetic field. They represent smooth
$1/B_{\perp}$-periodic oscillations of the PL-intensity at
spectral positions close to the Fermi level as the 2DES
transitions through conventional quantum Hall
states~\cite{APL2015}.

On further tilting of the field, we suddenly observe a
qualitatively different behavior: the spectral dynamics undergoes
abrupt transformations at magnetic field values close to filling
factors $\nu=$2, 4, 6,... . On the high-field side of this cutoff
($B^*$), the PL evolution coincides with the previous case of Fig.
1a, but at smaller fields, the spectrum undergoes a sudden
"reorganization": two narrow peaks emerge instead of three, and
their oscillator strengths become inverted (See top and bottom
spectra in Fig. 1d). The high-energy spectral line centered at
$\sim 3680.5$\,\AA \,\, for $\nu=2$ and also developed at other
even filling factors completely disappears (see Fig.1a). This
"reconstruction" of the PL specrum is indicative of the abrupt
magnetic-field-driven change in the ground state. The range of
these {\it critical} tilt angles supporting such transformations
is a few degrees, exceeding which the dynamics becomes "smooth"
again with the modified recombination spectrum close to even
filling factors (See panel (c) on Fig.1). Importantly, this new
appearance of spectra at even filling factors (the lowest spectrum
on Fig. 1d) is identical to those at $B<B^*$ near the cutoff
(Fig.1b) and thus represents the PL signature of the new phase of
the $\nu=2$ state. At tilt angles supporting abrupt PL
transformations, the 2DES undergoes a phase transition. As
discussed below and in accordance with previous magnetotransport
results~\cite{Stoner2012,FalsonFQHE2015}, the two phases around
$\nu=2$ correspond to PM ordering with opposite LL spin states
equally occupied and FM ordering with inverted spin orientation of
the highest occupied spin level. For the sake of brevity, the PL
spectra of different phases in Fig.1 are denoted as PM and FM. An
interesting interplay between the two phases can be observed
across the narrow B-field transition region. Here, the resulting
spectrum is composed of the superposition of the PM and FM spectra
(Fig.1d). The proportion between the phases is gradually switched
from the full PM to the full FM state over a span of $\Delta\,B <
0.1$\,T. An example of the best linear superposition fitting the
mixed PL spectrum is overlaid on the middle spectrum in Fig.1d,
and the relative weights are plotted in the inset therein. This
system state witnesses the co-existence of domains in a narrow
transition region of magnetic fields and facilitates direct
estimation of the phase percentage.

\begin{figure}[htb!]
\includegraphics[width=.48\textwidth]{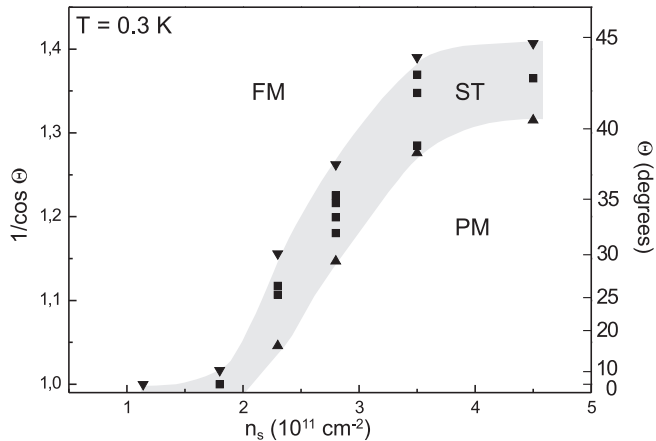}
\vspace{-2.mm} \caption{Phase diagram of Ising quantum Hall
ferromagnets at $\nu\approx 2$, obtained as a plot of the magnetic
field tilt angle versus electron density. Ranges corresponding to
ferromagnetic and paramagnetic phases and the Stoner instability
region are specified. Experimental symbols corresponding to each
of the samples mark the boundary tilt angles between different
regions (upward and downward triangles) and a few angles in
between corresponding to the Stoner transition. The shaded region,
encompassing the instability region, serves as a visual guide. For
the lowest density sample with $n_s=1.14\times
10^{11}$\,cm$^{-2}$, the paramagnetic phase at $\nu=2$ is absent,
and the Stoner ferromagnet is observed at all tilt angles.}
\label{fig3} \vspace{-1.5mm}
\end{figure}

Similar experiments were performed on all other samples listed in
Table \ref{table1} with electron densities ranging from
$1.14\times 10^{11}$ to $4.5\times 10^{11}$\,cm$^{-2}$. The data
are consolidated on a plot with critical angles as a function of
the electron density (Fig.2). Qualitatively similar behavior was
observed for all the 2DESs with densities exceeding $2\times
10^{11}$\,cm$^{-2}$, corresponding to three regions with different
magneto-PL dynamics around even filling factors. However, the
angles strongly depend on the electron density due to the
Fermi-liquid renormalization of the spin susceptibility. At n =
$1.8\times 10^{11}$\,cm$^{-2}$, the Stoner transition around
$\nu=2$ occurs even for the normal orientation of the magnetic
field. The instability region spans $\Theta\sim10^{\circ}$. At
even lower densities, a stable Stoner ferromagnet is formed even
at normal magnetic fields. Although the phase diagram corresponds
to the quantum Hall state $\nu=2$, these conditions perfectly
correlate with transformations at other even filling factors. In
addition, the level crossing with index $j=2$ (from
Eq.(\ref{Eq1})) was observed for $\nu=3$ for the sample with
density $2.8\times 10^{11}$\,cm$^{-2}$. The tilt angles for this
second coincidence lie in the very narrow range of $\Theta_2=63.5
\pm 0.5\,^{\circ}$, and this result is consistent with the results
of magnetotransport experiments for higher coincidence indices and
filling factors. This behavior may be attributed to less
pronounced exchange effects in level crossing.

An extended view of these phenomena was obtained by means of
inelastic light scattering experiments. Here, we expected the
probing of collective excitations to reveal the spin properties of
the ground state. As reported in the previous
study~\cite{RamanPRB2016}, inelastic light scattering signal
corresponding to neutral electronic excitations of 2DES can be
found in vicinity of its PL recombination lines. The point of
interest here is the estimation of appropriate resonance
conditions for each Raman spectral line and its further
identification. Data obtained in ~\cite{RamanPRB2016} for zero
magnetic field can be utilized to identify the set of intersubband
excitations at any given magnetic field simply by tracing the
resonances while continuously increasing magnetic field. In the
context of the Stoner transition, most actual are the excitations,
sensitive to the spin-degree of freedom. The direct sensor is an
intra-LL spin exciton (SE) -- collective mode, indicative of a
spin arrangement. Irrespective of the inter-particle correlations,
the energy of the SE is close to pure Zeeman splitting in the
long-wavelength case (Larmor's theorem~\cite{Larmor}). This
property makes the SE trivial for identification. In our
structures, this excitation appears in the Raman spectra over the
comparatively wide range of laser wavelengths of $\sim 9$\AA,
overlapping with the resonant conditions for an intra-LL
magnetoplasmon. Over a wide span of the magnetic field, the SE
energy increases linearly with an effective Lande factor
$g^*=2.00\pm 0.015$ (plot in Fig.3b). A feature of interest with
regard to this excitation is its most elementary structure with
just one quantum number (the total 2DES spin) changed by unity.
This hinders the decomposition of the SE into other elementary
excitations even when the system deviates from incompressible
states. For this reason, the SE is a "long-lived" excitation, and
its spectral width in the actual Raman spectra is determined
solely by the equipment resolution of $\sim0.2$\,meV, rather than
its thermal decay and inhomogeneous broadening. The typical SE
Raman spectrum is shown in Fig.3a.

\begin{figure}[htb!]
\includegraphics[width=.48\textwidth]{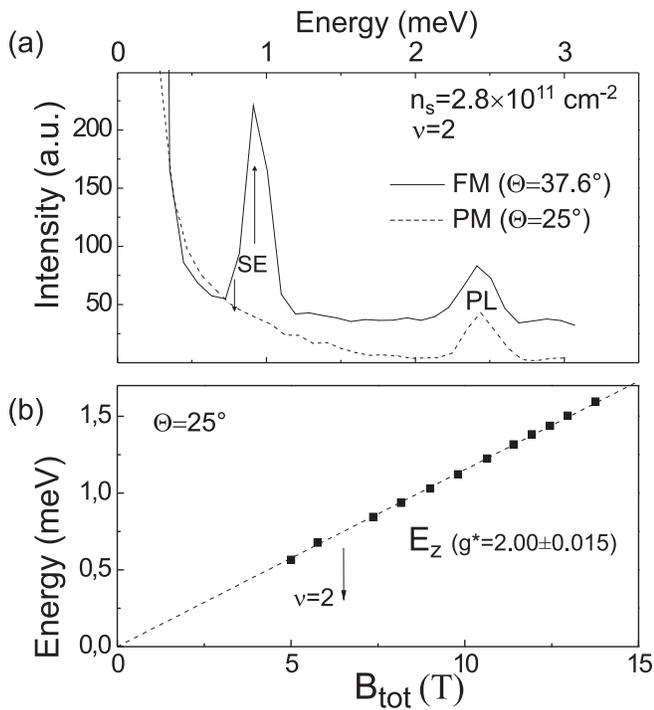}
\vspace{-2.mm} \caption{(a) Raman spectra of spin exciton at
$\nu=2$ and T = 0.3\,K at two different tilt angles, corresponding
to the ferromagnetic and paramagnetic phases (the SE is absent in
the PM phase). Arrows indicate the spectral positions of the SE
Raman line calculated for $g^*=2$. Apart from Raman signal, a
single photoluminescence line is present in the spectrum, which
can serve as a reference intensity level. (b) Energy of SE
measured as a function of the total magnetic field for sample
orientation of $\Theta=25^{\circ}$.} \label{fig3} \vspace{-1.5mm}
\end{figure}

The most indicative property of the SE is the change in its
spectral intensity with change in the spin ordering in the system.
According to the trivial spin-flip representation of SE, its
weight is obviously proportional to the number of occupied states
in the initial spin level and the number of vacancies in the final
level. Therefore, the SE should gain the maximum weight in systems
with FM ordering and conversely zero weight in the PM phases
corresponding to normal states with even filling factors. A
quantitative analysis of the SE spectral intensity behavior at
arbitrary filling factors is hardly possible, since it would
require knowledge of the microscopic structure of the ground state
and consideration of the orbital wave functions of participating
electrons. Nevertheless, the experimental dynamics of the SE line
intensity can serve as an indicator of the general asymmetry in
the spin-up and spin-down occupations and is particularly
meaningful in the vicinity of integer QH states.

Fig.4a depicts the magnetic field dynamics of the SE intensity in
the sample with density $n_s=2.8\times 10^{11}$\,cm$^{-2}$ at the
three angular orientations matching those of Fig.1. The horizontal
axis represents the normal component of the magnetic field such
that the positions of integer filling factors for all angles match
each other. We note (in Fig.4a, angle 25$^{\circ}$) that in the
ferromagnetic QH state at $\nu=1$, the intensity reaches a local
maximum as per the abovementioned considerations. Close to
$\nu=2$, the SE spectral weight reduces and abruptly drops to zero
over a certain range on both sides of $\nu=2$. Naturally, in this
PM phase with symmetric occupation of the spin-up and spin-down
states, the spin exciton is absent. Spontaneous symmetry breaking
occurs around $\nu=2$ at a higher angle, corresponding to the
Stoner instability region ($\Theta=32.1\,^{\circ}$ in Fig.4a).
Here, the intensity of the SE spectral line suddenly increases and
reaches a sharp maximum, thereby indicating the presence of
parallel spin alignment in the system. Interestingly, on the
right-hand-side region of $\nu=2$, the SE intensity, although it
drops dramatically, does not become zero, and therefore, the PM
order is not entirely "resumed" here. The position of this cutoff
in the SE behavior at angle $\Theta=32.1\,^{\circ}$ coincides with
the discontinuity observed in the magneto-PL dynamics and
evidently separates the two phases. The highest tilt angle for
this sample (37.6$^{\circ}$) corresponds to smooth dynamics both
in the PL spectra and Raman spectra of the SE line. A pronounced
local maximum in the SE intensity is observed right at $\nu=2$,
evidencing FM ordering of the 2DES. Apart from this feature, we
can also notice a qualitatively similar behavior of the SE on the
higher-magnetic-field side of $\nu=2$ at all angles. Taking this
observation into account, we conclude that system first undergoes
a spontaneous symmetry breaking close to $\nu=2$, which manifests
as a discontinuity in the energy spectrum at tilt angles in the
instability range. The spin ordering at fields slightly exceeding
the transition point deviates from paramagnetic and is not fully
compensated, probably due to the nucleation of the FM phase. For
angles lying outside this range, the Stoner transition does not
occur; instead, the whole neighborhood of $\nu=2$ corresponds to
the FM phase.

\begin{figure}[htb!]
\includegraphics[width=.48\textwidth]{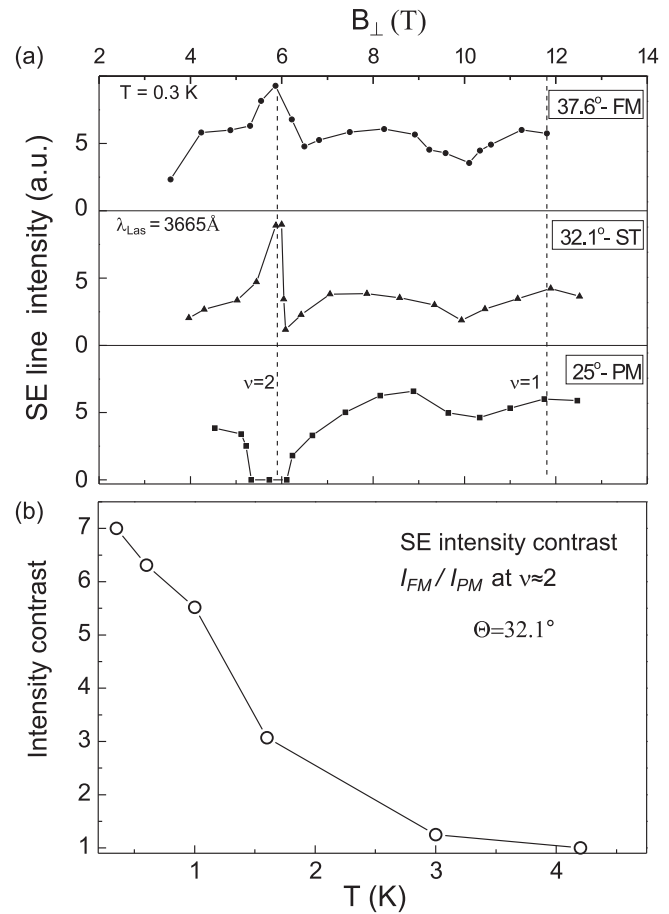}
\vspace{-2.mm} \caption{Behavior of intra-Landau-level spin
exciton spectral intensity as a function of the spin
transformations at $\nu\approx 2$ in the sample with $n_s=$
$2.8\times 10^{11}$\,cm$^{-2}$. (a) The magnetic field dependence
of the SE spectral intensity as recorded for three different tilt
angles. In the paramagnetic (PM) phase ($\Theta=25^{\circ}$), the
SE is absent near $\nu=2$, in the ferromagnetic (FM) phase
($\Theta=37.6^{\circ}$), its intensity exhibits a local maximum,
and at the Stoner transition ($\Theta=32.1^{\circ}$), the SE
behavior sharply switches between the PM and FM cases. In all
cases, the SE spectral intensity reaches maximum again in the QHF
state of $\nu=1$. (b) The temperature dependence of the SE line
intensity ratio on either side of the cutoff field, recorded at
angle $\Theta=32.1^{\circ}$.  } \label{fig4} \vspace{-1.5mm}
\end{figure}

The dramatic change in SE intensity across the Stoner transition
point can be considered as an indicator of a phase contrast
between PM and FM orderings. The thermal stability of these phases
was next probed in terms of the "smearing" of the SE intensity
collapse. At a fixed tilt angle and varying temperatures, the
intensity contrast $I_{FM}\,/\,I_{PM}$ was calculated as the ratio
of the SE intensities corresponding to the local maximum (FM
phase) and the local minimum (PM phase) near $\nu=2$. The dynamics
of this value was measured in the temperature range 0.3...4.2\,K
(see Fig.4b). From the figure, we note that the intensity contrast
drops with decrement corresponding to $\sim2$\,K. Effectively, the
temperature T=1.6\,K is the last temperature value at which the
Stoner-like discontinuity is observed. This result affords an
estimate of the Curie temperature for these QHF states, which was
also characterized in magnetotransport experiments as the limiting
temperature at which the domains of the two phases exist in the
system~\cite{AlAs2003}.

\begin{figure}[htb!]
\includegraphics[width=.48\textwidth]{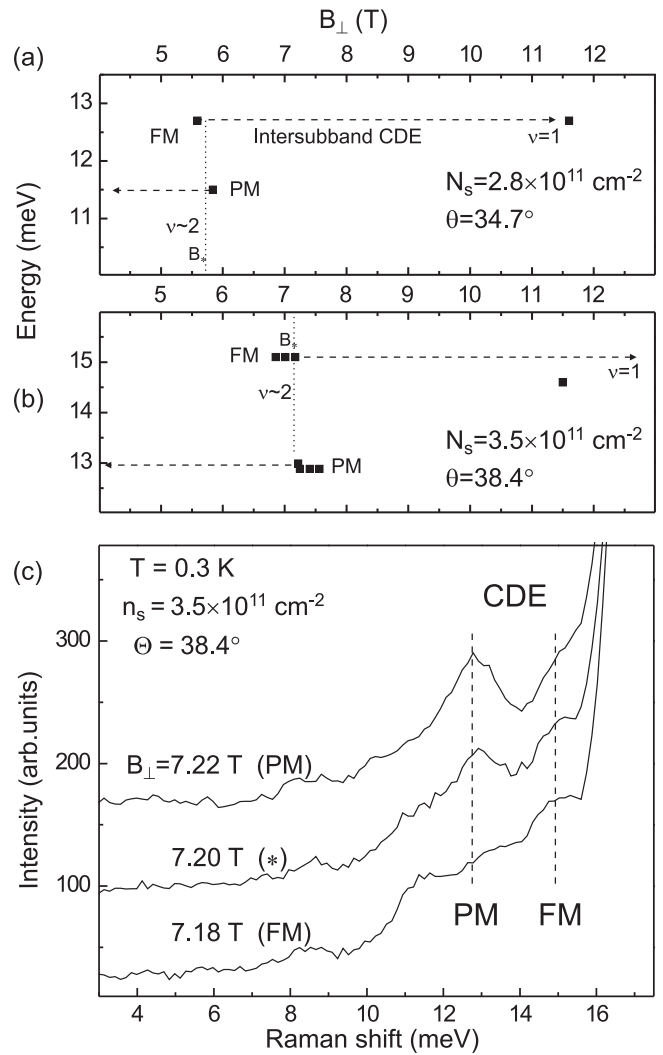}
\vspace{-2.mm} \caption{(a) and (b) Energies of intersubband
charge density excitation (CDE) measured for two different samples
at tilt angles supporting Stoner transition. The sharp energy
shift at $\nu\approx 2$ is due to spin rearrangement at the
ferromagnetic-paramagnetic (FM-PM) phase transition. Horizontal
dashed lines are intended to align CDE energies on either sides of
$\nu=2$ to those of spin-polarized ($\nu=1$) and spin-unpolarized
(zero magnetic field) states. (c) Raman spectra of intersubband
CDE in the sample with $n_s=3.5\times 10^{11}$\,cm$^{-2}$ at three
closely lying magnetic field values around the Stoner transition
point. The CDE lines at two energy positions are interplayed in
intensity according to the phase proportion. Their positions are
marked with verical dashed lines.} \label{fig5} \vspace{-1.5mm}
\end{figure}

An additional tool sensitive to local spin configuration is
represented by the energy of the long-wavelength collective
excitations. Recently, the intersubband plasmon (or charge density
excitation - CDE) has been shown to reflect the spin-polarization
degree owing to changes in the exchange energy
contribution~\cite{Kulik2017}. Being not particularly susceptible
to LL occupation, this intersubband mode nevertheless exhibits a
qualitatively different structure between FM and PM spin
orderings. A considerable shift in the CDE energy ($\sim$1 to 2\,
meV) has been detected in several ZnO-based structures while
continuously tuning the system from an unpolarized $\nu=2$ state
to the FM state $\nu=1$. In view of our study, the energy shifts
of the intersubband CDE at $\nu=2$ behave in accordance with the
spin transformations. First of all, the CDE energy in the $\nu=2$
PM phase at small tilt angles is equal to that at B\,=\,0\,T or
high LL filling factors. An identical CDE spectrum is observed on
the high-field-side of the cutoff point at tilt angles in the
instability range (see the topmost spectrum in Fig.5c and
lower-energy data symbols in Figs.5a and 5b). This is clear
evidence that the spin-unpolarized states possess intersubband CDE
with equal energies. The abrupt shift in the CDE energy occurs
while tuning the magnetic field across the cutoff point (the
lowest spectrum in Fig.5c). This spectrum is identical with that
in the neighborhood of $\nu=2$ at tilt angles lying in the FM
region, and most interestingly, it is similar to that of the FM
state at $\nu=1$. This situation is illustrated in Figs.5a and 5b,
corresponding to two samples with different densities. The CDE
energies on both sides of the Stoner transition are aligned to
those at the FM $\nu=1$ state (where it is acessible by the
solenoid) and B=0\,T. Furthermore, the two CDE lines are observed
simultaneously in the narrow range of fields across the Stoner
transition (the middle spectrum in Fig.5c). This means that a
surface-integrated Raman signal contains the linear superposition
of separate spectra from the two phases. This information is
valuable since the presence of unperturbed energies of the
collective excitations, emanating from the domains of different
phases, is possible provided their sizes exceed at least few
magnetic lengths. In general, this conclusion agrees with
theoretical expectations based on the Ising
model~\cite{MacDonald2001}.

\section{Discussion}

The control of the spin configuration at integer filling factors
by means of magnetic field tilting was shown to differ from the
single-particle considerations. A wide range of tilt angles is
suitable for affording a level coincidence at even filling
factors. This implies a substantial exchange contribution in the
coincidence condition. For this reason, the phase diagram of QHFs
in terms of critical tilt angles at even filling factors contains
three essential regions: stable PM phase with anti-parallel spin
orientation, stable FM phase with a pronounced maximum of spin
polarization at even filling factors, and an instability region
supporting a phase transition near even filling factors.
Fermi-liquid effects, dependent on electron density, cause
substantial shift of the critical tilt angles. At densities lower
than $\sim2\times 10^{11}$\,cm$^{-2}$, the stable PM phase at
$\nu=2$ is no longer accessible, and instead, a Stoner transition
occurs at small angles. At even lower densities of
$\sim10^{11}$\,cm$^{-2}$, the FM phases are stable independently
of the tilt for at least several even filling factors
$\nu=2,4,...$.

These sharp transformations of the 2DES phases are perticularly
amazing, bearing in mind the substantial macroscopic nonuniformity
in the electron density, which is estimated to be $\delta
n_s\sim$\,5-10$\%$, depending on the sample. The Stoner transition
appears to be governed by a certain coherent process, aligning a
chemical potential level throughout the entire 2DES. Therefore,
from the plot in the inset in Fig.1d, we note that the entire
transition fits in less than $\delta B\sim0.1$\,T, within
$\sim1.5\,\%$ of the whole field. The range of fields, extracted
from the SE intensity drop, correlates with that of the PL
transformation. The proportion of the two phases is most
effectively extracted from the PL-spectra interplay, since the PM
and FM phases have specific signatures and in addition provide a
strong signal. This approach can also aid in characterizing the
subtle hysteretic behavior of the Stoner transition at $T=0.3$\,K.
So, the sweep-direction difference in the cutoff positions was of
the order 0.01\,T for sample 427 with density in the middle of the
studied range. This is likely due to the easy-moving domain walls
in low-disorder systems.

The non-vanishing SE Raman line on the higher-field side of the
transition region corresponds to residual asymmetry in spin
occupation despite signs of a dominant PM phase. This is most
probably due to the nucleation of the FM phase, which is however
too sparse to be discernible in the PL-spectra distortion or
modification of the CDE mode. This fact contradicts the existing
hypothesis concerning pure spin states on either side of the
Stoner transition point~\cite{IsingPolish2002, FalsonFQHE2015}.
The SE is particularly sensitive to perturbations of
spin-unpolarized states, since its intensity is better traced on a
zero background. The symmetrical situation with nucleation of the
minority phase is possible at fields in the FM region. However the
SE intensity can hardly serve for detection of this, since
intensity deviations should have been counted from an unknown
maximum intensity level.

The SE line at other even filling factors ($\nu=4,6,...$) flashes
in phase with ferromagnetic order, although its visibility is
poorer than that of the FM state $\nu=2$ due to the stronger
laser-line background. A qualitatively similar response was
obtained for the phase transition between two spin configurations
at $\nu=3$. In this case, for coincidence angle
$\Theta_2=63.5\,^{\circ}$, the SE intensity drops by a factor of
$\sim2.4$ across the cutoff point. This value reasonably agrees
with naive single-particle picture of spin-state occupation at
this filling factor which corresponds to $I_{max}\,/\,I_{min}=3$.

The thermal degradation of the SE contrast at a Stoner transition
qualitatively shows that domains with rivaling phases undergo
melting at temperatures of $\sim$2\,K, which value is
significantly lower than Zeeman and exchange energies, but it
appears to be comparable with the Curie temperature estimated for
an easy-axis QHF in terms of domain-wall
excitations~\cite{MacDonald2001} $T_C\sim 0.009\,e^2/\varepsilon
\ell_B$. This calculation as applied to sample 427 at $\nu=2$,
$B_{\perp}=5.9$\,T, and $\varepsilon=8.5$ yields $T_C\sim 1.6$\,K.
It is also worth noting that at temperatures above 4\,K, the
behavior of the SE intensity as a function of the magnetic field
is absolutely flat, with no peculiarities at integer QH states.

Concerning the domain size, the SE line is of no utility since it
reflects a general asymmetry in spin occupation. More indicative
of the domain size are the energies of the collective modes. In
our study, the observed intersubband CDE energy shift indicates
that at a transition point, the macroscopic QH domains of
different phases coexist and that their sizes exceed few magnetic
lengths, since a collective mode has unperturbed energy and
spectral width. According to theoretical
expectations~\cite{MacDonald2001},  at a transition point
(resistance spike maximum in magnetotransport), the typical domain
sizes are about dozens of magnetic lengths, but on deviation from
the transition point by an effective magnetic field of $\delta
B\sim 0.02$\,T, the minority phase domains shrink to a size of
$\sim3\,\ell_B$. This result is consistent with our conclusion,
since at small deviations from the central field of the Stoner
transition, the minority phase CDE line vanishes (Fig.4c).

\section{Conclusion}

In conclusion, we performed a magneto-optical study of the Stoner
transition between quantum Hall ferromagnetic states in 2D
electron systems based in MgZnO/ZnO heterostructures. Abrupt
transformations were detected both in the photoluminescence
spectra and in 2D collective excitations, probed by resonant
inelastic light scattering. The transition was facilitated by the
tilting of the magnetic field, thus bringing spin-Landau levels to
coincidence. Depending on the 2D electron density, the spin
configuration varied at different critical angles. This phenomenon
was exploited to draw the phase diagram for quantum Hall
ferromagnets at $\nu\sim 2$ in terms of the tilt angle vs 2D
electron density. Three qualitatively different regions were
identified: a stable paramagnetic phase around even filling
factors, a stable ferromagnetic phase, and a Stoner instability
region in between. At low electron densities of
$\sim1-2\times10^{11}$\,cm$^{-2}$, Fermi-liquid effects lead to a
Stoner transition even at normal magnetic field orientation. The
spin configuration at each state was characterized via the
spectral weight of the intra-Landau-level spin exciton. The spin
exciton intensity displays a maximum-like behavior in the vicinity
of the ferromagnetic states. At a Stoner transition point near
$\nu=2$, it switches from a sharp maximum (in the ferromagnetic
phase) to a deep minimum (paramagnetic phase). However, it does
not vanish completely, thereby indicating the presence of some
ferromagnetic nucleation in the paramagnetic state. From thermal
smearing of the spin exciton dynamics, the Curie temperature of
quantum Hall ferromagnets was estimated to be $\sim2$\,K. Optical
signals from the domains were observed to superimpose, which
facilitated the direct calculation of the phase proportions across
the transition point. It was also shown that intersubband
collective excitations from the two different phases coexist at
the transition point and are unperturbed, thus indicating that the
domain sizes exceed few magnetic lengths.

\begin{acknowledgments}
We acknowledge financial support from the Russian Scientific
Foundation (Grant No.14-12-00693).
\end{acknowledgments}


\begin{thebibliography}{99}


\bibitem{FalsonGrowth}
J. Falson, Y. Kozuka, J. H. Smet, T. Arima, A. Tsukazaki and M.
Kawasaki, Applied Physics Letters {\bf 107}, 082102 (2015).

\bibitem{ReviewZnO}
Y. Kozuka, A. Tsukazaki, and M. Kawasaki, Applied Physics Reviews,
{\bf 1}, 011303 (2014).

\bibitem{Stoner2008}
A. Tsukazaki, A. Ohtomo, M. Kawasaki, et al., Phys.Rev.B {\bf 78},
233308 (2008).

\bibitem{Stoner2012}
Y. Kozuka, A. Tsukazaki, D. Maryenko, J. Falson, C. Bell, et al.,
Phys.Rev.B 85, 075302 (2012).

\bibitem{Maryenko2014}
D. Maryenko, J. Falson, Y. Kozuka, A. Tsukazaki, and M. Kawasaki,
Phys.Rev.B {\bf 90}, 245303 (2014).


\bibitem{FalsonFQHE2015}
J. Falson, D. Maryenko, B. Friess, D. Zhang, Y. Kozuka, A.
Tsukazaki, J. H. Smet and M. Kawasaki, Nature Physics 11, 347
(2015).



\bibitem{AlAs2000} E. P. De Poortere, E. Tutuc, S. J. Papadakis, M. Shayegan, Science {\bf 290}, 1546 (2000).

\bibitem{AlAs2003} E. P. De Poortere, E. Tutuc, and M. Shayegan,
Phys.Rev.Lett. {\bf 91}, 216802 (2003).

\bibitem{IsingPolish2002} J. Jaroszyn´ ski, T. Andrearczyk, G. Karczewski, et al., Phys.Rev.Lett. 89, 266802 (2002).

\bibitem{IsingInSb} J. C. Chokomakoua, N. Goel, S. J. Chung et al., Phys.Rev. B, 69, 235315 (2004).

\bibitem{MacDonald2001} T. Jungwirth and A. H. MacDonald, Phys.Rev.Lett. {\bf 87}, 216801 (2001).


\bibitem{APL2015} V. V. Solovyev, A. B. Van'kov, I. V. Kukushkin, et al. ,
Appl.Phys.Lett. {\bf 106}, 082102 (2015).

\bibitem{RamanPRB2016}
A. B. Van'kov, B. D. Kaysin, V. E. Kirpichev, V. V. Solovyev, and
I. V. Kukushkin, Phys.Rev. B {\bf 94}, 155204 (2016).

\bibitem{Larmor}
M. Dobers, K. von Klitzing, and G. Weimann, Phys. Rev. B {\bf 38},
5453 (1988).


\bibitem{Kulik2017}
L.V. Kulik, A.B. Van'kov, B.D.Kaysin, and I.V. Kukushkin, JETP
Letters, {\bf 105}, i.6, p. 358 (2017).



\end{thebibliography}
\end{document}